# Collapse of cooperation and corruption in a mathematical model within game theory including Moldovan case study (Homo Sociologicus vs Homo Economicus)

*Jarynowski, Andrzej*
*Moldova State University/Jagiellonian University, andrzej.jarynowski@sociology.su.se*

**Resume**: A simple model from game theory, which can imitate a mechanism of corruption and cooperation patterns, is proposed. The settings are divided into two studies with examples related to safety of Moldovan economy. In Homo Economicus world, players seems to act in rational way and decisions are driven by payouts (metaphor for behavior in populations of selfish agents). In Homo Sociologicus world decisions are described by players acquiring reputation and evolving altruism, which in turn determine their choice of strategy (evolution of cooperation). I shall discuss both modeling approaches of Prisoner Dilemma, to understand collapse of cooperation and increase of corruption in post-communistic countries as Republic of Moldova. Possible more efficient for society outcomes (fairness equilibrium) are also discussed.

**Introduction to game theory (the prisoner's dilemma).** Game theory problems are well studied in economics and individual human decision can be understood this way. The authorship is attributed to John von Neumann in 1928 and later solution to John Nash in 1950, although the parallel issue was described by Hugo Steinhaus: mathematician associated with Wroclaw and Lviv (Stainhaus, 1963). Cooperate or defect: these questions can be answered according to some mathematical rules. There are many types of games from single tot the repetitive one (which can example adaptable strategy - based on history). An interesting case of the problem initially considered by physicists is a minority game. The payoff decreases drastically when too many players choose the same strategy. However, I concentrate the basic game in game theory: the prisoner's dilemma. According to the classical theory, the mutual cooperation is never optimal from an individual point of view, but always profitable for society. Let's explain Prisoner Dilemma in details. Two criminals were catch by the police. If the prisoner 1 confesses (defeat by betraying his colleague) and prisoner 2 deny to accuse prisoner (cooperate - will try to fair to his colleague). So 1 will be a prisoner acted as a witness against second and go to prison for a short time ($T$ years), while the prisoner 2 will get full judgment ($S$ years) of sentence (and vice versa). If both deny (cooperate) and nobody confesses, they will be convicted ($R$=0 years). If both defeat (betray each others), both receive the same average sentence ($P$ years). The order of payments in such a game would be in order from the most profitable: $R$=0 years <$T$ years <$P$ years <$S$ years. Such a system has an equilibrium point where cheating is the most optimal strategy. Many other problem could be described in such a setup and in paper players are the agent on economic market and games represents transactions. I will focus on optimization of "social" goals, while agent pursue only their own material self-interest with exception to altruism.

**Table 1** Payout matrix of standard Prisoner Dilemma

| Players ½ | C (cooperation) | D (defection) |
|---|---|---|
| C (cooperation) | R, R | T, S |
| D (defection) | S, T | P, P |

The analysis below is only indicative, but it will still give only insights to politician economists in scope of the dangers within national Economy for such a state like Republic of Moldova.

**Economics.** I would like to model non-competitive market, where oligopoly and corruption could appear and agents (players) are: companies, state bodies, etc. As markets and economy are made up of people reacting to what others are doing (as in Prisoner's Dilemma). One's actions affect second as wells as everyone else's welfare within given Economy. Achieving a cooperative solution is very difficult if there is a change for exploitation others players or state (Keynes, 1936). Let's try understand interactions in national economy where in a single game, the agents can choose two strategies: to be the donor (Do) or acceptor (A) of bribe. If both agents always want to be a bribe acceptor, it leads to no illegal transaction (this case would describe a fair economy). The outcome (payoff) of such a game would is based on the fair market competition or cooperation (which is in-differentiable). If both agents choose Do, illegal transaction do not occurs between agents. If both agents want to give a bribe, it can lead to a situation where one buy some advantages from third site e.g. government official (for simplicity, let us assume that only the agent supporting ruling political party can bribe the government official). We must unify units of factors for different payoffs: the ability to compete in given game for both agents (depending on the position and adaptability of agents on the market), the amount of monetary/valued bribes, profit or loss in effect of corruption, the moral cost of participation in corruption and expected value of punishment (if caught).

**Table 2** Payout matrix of corruption system

| ½ | Do | A |
|---|---|---|
| Do | (W1-L, W2+G-C- m)* | W1+G-C- m, W2-L+C- m |
| A | W1-L+C- m, W2+G-C- m | W1, W2 |

Expanding symbols (2): W1 and W2- potentials of players ("market values" of them), G- gain due to illegal transition L- lost due to illegal transition, C - value of a bribe, m - negative aspects of illegal transaction as sum of moral costs $m_C$ and expected value E(X) of fine cost F relative probability ($p_{Catched}$): m= $m_C$ + E(X)= $m_C$ + (1- $p_{Catched}$)*0+ $p_{Catched}$ *F. Star in Table 2(*) corresponds to both donors game (DoDo), where state authority is taking a bribe.

From the long perspective of economy, the model parameters should depend on time or history and it would correspond to the evolutionary thunder. The provided analysis is only indicative, but the most important factor driving the corruption is a profit function: $G - C - m$. To do a proper economic analysis, we must bring all the factors to utility (as in classical and neoclassical economics methodology). Afterward, one could look for equilibria (Nash-type or Min-Max) and compare the resulting strategies.

We note that the game theory has already came several times to describe the phenomenon of corruption. Understanding mechanisms of corruption is a step to combat this practice (Jarynowski, 2010). In countries where corruption is rife as Republic of Moldova, there are suggestions on reducing regulation (transparency) by state authority of firm or individual behavior in an effort to ensure that private and public interests are brought into line (move attractiveness from strategy DoDo to AA and decreasing factor G). Effective legal sanctions can change the outcome of the corruption game, namely, decision of the parties to engage in a corrupt transaction. It leads to a new equilibrium with lower corruption level, but never allows to rid off this problem. Moreover, implement public fines or other sanctions (increasing probability of catching - $p_{Catched}$ and fines value - F) is not optimal if these agents (state authorities) are already engaged in taking bribes themselves (Macrae, 1982).

Let illustrate it on example of possible case in Moldova (based on private correspondence of the author). Two foreign investors want to buy sugar factory. A large proportion of the times and the energy of entrepreneurs and businessman is spent dealing with government agencies in developing countries. Clients prefer bribing, instead of waiting long queues or getting rejected their applications. Rules, regulations, state monopolies, rationing of goods that are short in supply and publicly owned firms create many opportunities for corruption (countries with liberal economy and free market experience less corruption). Excessive bureaucracy (even due to anti-corruption regulation) or red carpet (huge competence of state actors) cause usually more bribe to government officials (Bayar, 2003). Both of mentioned investors hired local - Moldovan consultants (intermediates) with "extra allowance" to proceed transaction (corruption acts as an implicit tax). Let assume consultant of investor 2 has connections to ruling party. Its representation of DoDo scenario and corresponding to Table 2, gain of this treatment must be greater than costs (G>C+m). Then, Player 2 is the winner and player 1 is the looser, but society is also on lost position (-G as inefficiency in the market). Moreover, cost of preparation to investment (L) is lost to player 1 and probably this agent will not invest in Moldova any more (discourages of entrepreneur, which in turn affects investment, growth and development of the country).

**Sociology.** Building a cooperative society involves moral education; values, norms of the society. Post-communism transformations leaded to protection of political elites with implementation democratic procedures, with huge mistrust levels among citizens. Private networks are very often engine of economy and cooperation is missing due to disappointment of previous as current institutional system. Lack of cooperation leads to severe economic difficulties, massive corruption, and make society poorer and unequal (Howard, 2003). However, human societies are organized around altruistic, cooperative interactions. The emergence of moral/social norms and rebuild the trust, even destroyed during communism time, is still possible (Nowak, 2005). Let consider model of game where agents plays in pairs and have two possibilities: to cooperate (C) or to defect (D). The probability of cooperation depends linearly, both on the player's altruism and the co-player's reputation. Collective behavior is introduced by altruistic optimism (punishment) and reputation reciprocity (fail). Agents can establish the best strategy in repeated games. The final, stationary probability of cooperation can vary sharply with the initial conditions and jumps to minimal or maximal value for some critical conditions. Specification of the rules as initial conditions have impact on final states as well as on dynamics of the system. If only the reputation could vary one would observe coexisting strategies [Fig. 1] but with altruism change all players choose only one strategy [Fig. 2]. In both approaches, payoffs are not relevant and only mutual interaction between players are significant. We also observe, that the transition state close to the boarder between the two regimes can be described as Gaussian cumulative distribution function.

Each of two identical players has two different strategies: to cooperate (C) with the other or to defect (D) from cooperation. The probability that $i$ cooperates with $j$ depends on level of (Jarynowski, et al. 2012):
- Reputation $W$ of co-player
- Altruism $\varepsilon$ of player $i$, as a measure of her/his willingness to cooperate with others.

$$P(i,j) = W_j(i) + \varepsilon_i$$

If $P(i,j)>1$ is set 1, if $P(i,j)<0$ then 0. Reputation - $W$ is in range of [0,1] and Altruism - $\varepsilon$ is in range of [-1/2,1/2]. Reputation of player increase (decrease) if he cooperates (defects). Altruism of player increase (decrease) if co-player cooperates (defects). Speed of change for reputation/altruism is defined by $x_W/x_\varepsilon$ as a percentage change of reputation / altruism.

(C) $\varepsilon := (0,5-\varepsilon)x_\varepsilon + \varepsilon$
(D) $\varepsilon := \varepsilon + (-0,5-\varepsilon)x_\varepsilon$
(C) $W := (1-W)x_W + W$
(D) $W := W - Wx_W$

Possible mutual choices (in agreement with Prison Dilemma's notation from Table 1) : R- both cooperate, S- co-player defects when a player has cooperated, T- player defects when the co-player has cooperated, P- both defect.

In our simulations, there are from 100 to 1000 players in game with some initial conditions $W$ and $\varepsilon$. Connections between players are implemented as a fully connected graph, square lattice or Erdős–Rényi graph (where private networks of contacts seem to be very important).

**Table 3** Model scenarios and outcomes for different configurations

| Parameters/ scenario | Model descriptions | Observations |
|---|---|---|
| $x_W$=0,5 and $\varepsilon$=const | Reputation changes in „bisection" way and altruism is constant | Mutual choices exist and create symmetric coexisting frequencies curves [Fig. 1 left] |
| $x_W$=W' and $\varepsilon$=const | W' is reputation of co-player, so player's reputation change as quickly as W' | Symmetry broken – cooperation is more common, unstable final states [Fig. 1 right] |

| $x_W$=0,5 and $x_\varepsilon$=0,5 | Reputation and altruism changes in „bisection" way the same time | All players choose only one strategy, and the initial state is divided into two attractors [Fig. 2] |
|---|---|---|
| $x_W$=0,5 and $x_\varepsilon$={0,5;0} | Altruism changes only in some cases: CC (goes up) and CD (goes down) | Symmetry broken – cooperation is more common, non-full negative state |
| $W_i(i)$ | Individual vision of agent's W | System slows down |
| E-R or lattice | Agents on special networks | Spatial correlations appear |

Firstly, let consider constant altruism scenarios $\varepsilon$=const ($x_\varepsilon$=0). If altruism is unchangeable lifelong, all spectrum of behaviors (R S, T, P) are observed in the system [Fig. 1 left].

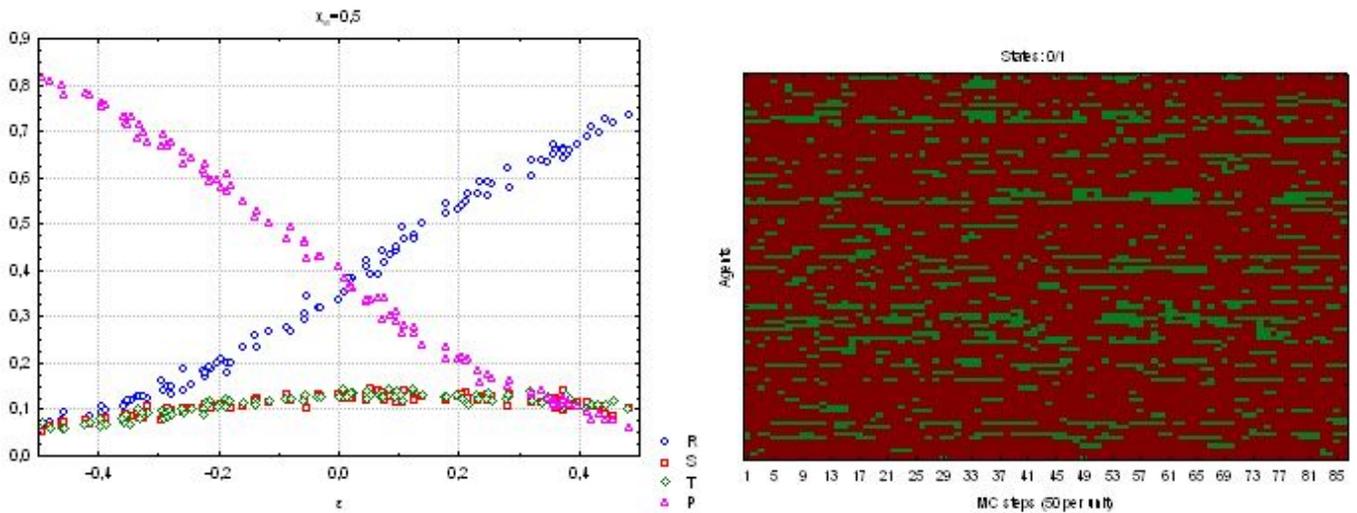

**Fig. 1** $\varepsilon$=const. Left - Frequencies of mutual choices for $x_W$=0,5. Right - Spontaneous transitions between reputation's states (0/1) for $x_W$= W' („Positive" state 1 dominates).

Let consider case where both $x_W$ and $x_\varepsilon$ are non-zero. For simplicity, let assume that both are equal to 0,5. Because of both parameters change, we could draw strategies versus constant vector of initial states $W$ and $\varepsilon$.

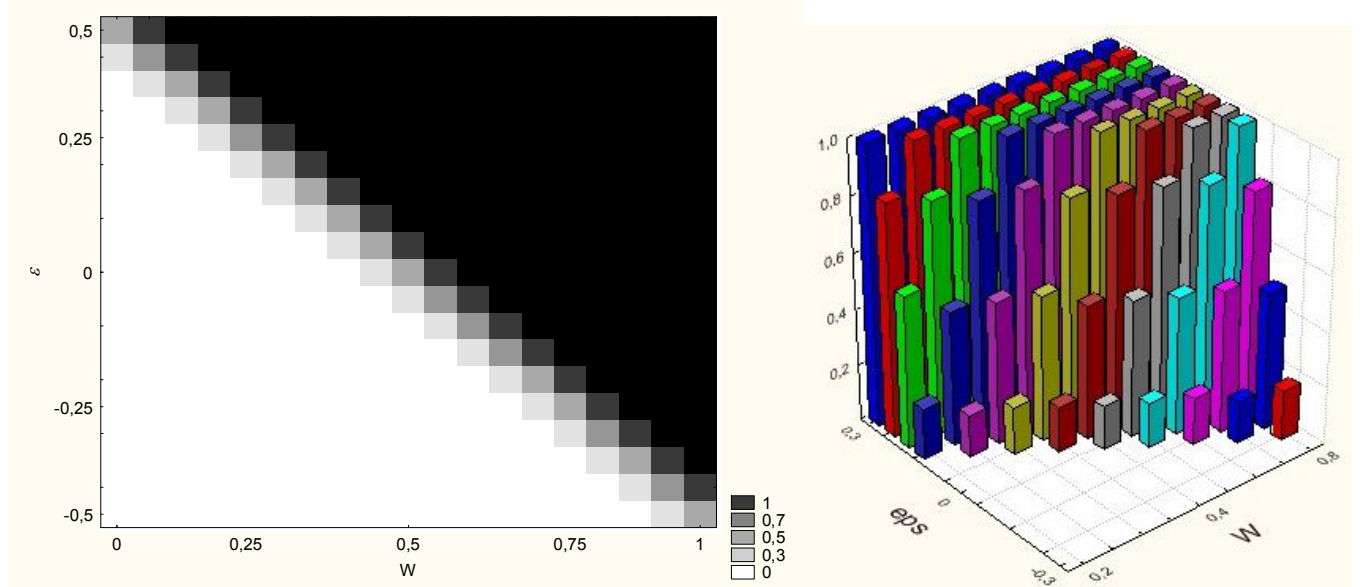

**Fig. 2** Phase separation of clean cooperating or defecting for given initial conditions, the mixed phase around anti-diagonal

Once the altruism is allowed to evolve, in long time limit the simulated players adopt one strategy R or P, the same for the whole population [Fig. 2]. Direction of the process is closer to real situation rather, than the stationary state in the long time limit. Sociotechnics and system controlling is possible. If symmetry of final state is broken – cooperation state is promoted (social norm works).

Based on observation of many of post communistic countries, Moldovan economy is dominated by defecting. If it would be possible to convince enough amount of companies and state actors to cooperate, then reputation (and even altruism as a derivative of it) will also growth, which make more people cooperating (feedback loop). In $x_W$=0,5 and $\varepsilon$=const scenario, if level of cooperation increase after some time system will come back (relaxation) to its equilibrium state [Fig. 1 left]. In network topology, such a increase of cooperation would be observed only locally and will also relax. However, in $x_W$=0,5 and $x_\varepsilon$=0,5 scenario, a big enough temporal increase of cooperation will lead to depolarization of whole system and all agent would only cooperate. In other examples of conditional evolving altruism increase of cooperation would lead to new, more cooperative stable state. Even that nobody knows, which scenario is closer to reality, for most of them even temporal change of people behavior could lead to long-term effects like rebuilt of trust – the motor of economic development (Weber, 1922).

**Conclusions.** Bureaucratic rules, permits, licenses lack of transparency and mistrust in others lead to the occurrence of defecting and corruption. Analytical mathematical derivation and computer simulations describe what agents with different cognitive capabilities are likely to do in market system. This is a branch of behavioral economics with usage of sociological regularity applied rationality assumptions (Camerer, 2003). Economic (rational) and sociological (normative) approaches shows threats as well as opportunities for economy of post communistic countries. My models show which parameters are most important in observed phenomenon. From economic perspective, I propose to focus on decreasing profits (payoffs) of corruption by decreasing function of gain *G* and increasing function of potential costs *m*. From sociological perspective, I would recommend rebuilding social norms, by promoting cooperative attitude between the most susceptible (youth generation). Lack of cooperation or fair competition must be further analyzed from the viewpoint of social costs it caused. I use game theory to figure out what it is likely to happen in a strategic interaction (or chains of interaction) of persons or companies from society perspective and ask how to improve an existing situation.